\documentclass[prc,preprint,showpacs,showkeys,nofootinbib]{revtex4}%
\usepackage{graphicx}
\usepackage{bm}
\usepackage{epsfig}
\usepackage{amsmath}
\usepackage{amsfonts}
\usepackage{amssymb}%
\begin{document}
\title{RADCAP: a potential model tool for direct capture reactions}
\author{C.A. Bertulani}
\affiliation{National Superconducting Cyclotron Laboratory,
Michigan State University, East Lansing, MI\ 48824, USA}
\email{bertulani@nscl.msu.edu}
\date{\today}

\begin{abstract}
A computer program is presented aiming at the calculation of bound and
continuum states, reduced transition probabilities, phase-shifts,
photo-disintegration cross sections, radiative capture cross sections, and
astrophysical S-factors, for a two-body nuclear system. The code is based on a
potential model of a Woods-Saxon, a Gaussian, or a M3Y, type. It can be used
to calculate nuclear reaction rates in numerous astrophysical scenarios.

\end{abstract}
\pacs{25.40.Lw,25.20.-x, 26}
\keywords{Astrophysical S-factor, photo-disintegration, potential model}
\maketitle

\address{National Superconducting Cyclotron Laboratory, Michigan State University,
E. Lansing, MI 48824}

\draft

\narrowtext

{\Large PROGRAM SUMMARY}

\begin{enumerate}
\item \textit{Title of program: }RADCAP (RADiative CApture)

\textit{Computers:} The code has been created on an IBM-PC, but also runs on
UNIX machines.

\textit{Operating systems:} WINDOWS or UNIX

\textit{Program language used:} Fortran-77

\textit{Memory required to execute with typical data:} 8 Mbytes of RAM memory
and 2 MB of hard disk space

\textit{No. of bits in a word: }32 or\textit{ }64

\textit{Memory required for test run with typical data:} 2 MB

\textit{No. of lines in distributed program, including test data, etc.:} 3054

\textit{Distribution format:} ASCII

\textit{Keywords:} Potential model; Photodissociation; Radiative capture;
Astrophysical S-factors

\textit{Nature of physical problem:} The program calculates bound and
continuum wavefunctions, phase-shifts and resonance widths, astrophysical
S-factors, and other quantities of interest for direct capture reactions.

\textit{Method of solution:} Solves the radial Schr\"{o}dinger
equation for bound and for continuum states. First the eigenenergy
is estimated by using the WKB method. Then, a Numerov integration
is used outwardly and inwardly and a matching at the nuclear
surface is done to obtain the energy and the bound state
wavefunction with good accuracy. The continuum states are obtained
by a Runge-Kutta integration, matching the Coulomb wavefunctions
at large distances outside the range of the nuclear potential.

\textit{Typical running time:} Almost all the CPU time is consumed by the
solution of the radial Schr\"{o}dinger equation. It is about 1 min on a 1GHz
Intel P4-processor machine for a Woods-Saxon potential.
\end{enumerate}

\pagebreak

{\Large LONG WRITE-UP}

\section{Introduction}

In astrophysically relevant nuclear reactions two opposite reaction mechanisms
are of importance, compound--nucleus formation and direct reactions (for more
details, see, e.g., \cite{Rol88}). At the low reaction energies occurring in
primordial and stellar nucleosynthesis the direct mechanism cannot be
neglected and can even be dominant. The reason for this behavior is that only
a few levels exist for low excitations of the compound nucleus.

In order to calculate the direct capture cross sections one needs
to solve the many body problem for the bound and continuum states
of relevance for the capture process. There are several levels of
difficulty in attacking this problem. The simplest solution is
based on a potential model to obtain single-particle energies and
wavefunctions. In numerous situations this solution is good enough
to obtain the cross sections within the accuracy required to
reproduce the experiments.

In this article a computer program is described which aims at calculating
direct capture cross sections, based on a potential model. The program
calculates bound and continuum wavefunctions, phase-shifts, energy location of
resonances, as well as the particle-decay width, photodisintegration cross
sections, radiative capture cross sections and astrophysical S-factors. The
formalism for this model has been developed in Refs. \cite{CD61,TP63,RO73}.

\section{Bound states}

The computer code RADCAP calculates various quantities of interest for
two-body fusion reactions of the type%
\begin{equation}
a+b\longrightarrow c+\gamma,\ \ \ \ \mathrm{or}\ \ \ \ a\left(  b,\gamma
\right)  c\ . \label{apb}%
\end{equation}
The internal structure of the nuclei $a$ and $b$ is not taken into
account. Thus, the states of the nucleus $c$ is obtained by the
solution of the Schr\"{o}dinger equation for the relative motion
of $a$ and $b$ in a nuclear + Coulomb potential. Particles $a$,
$b$, and $c$ have intrinsic spins labelled by $I_{a}$, $I_{b}$ and
$J$, respectively. The corresponding magnetic substates are
labelled by $M_{a}$, $M_{b}$ and $M$. The orbital angular momentum
for the relative motion of $a+b$ is described by $l$ and $m$. In
most situations of interest, the particle $b$ is a nucleon and $a$
is a \textquotedblleft core\textquotedblright\ nucleus. Thus it is
convenient to
couple angular momenta as $\mathbf{l+I}_{b}\mathbf{=j}$ and $\mathbf{j+I}%
_{a}\mathbf{=J}$, where $\mathbf{J}$ is called the channel spin.
Below we also use the notation $\mathbf{s}$, instead of
$\mathbf{I}_{b},$ for the intrinsic spin of particle $b$.

The bound state wavefunctions of $c$ are specified by
\begin{equation}
\Psi_{JM}\left(  \mathbf{r}\right)  =\frac{u_{lj}^{J}\left(  r\right)  }%
{r}\mathcal{Y}_{JM}^{l}\ , \label{psi_expansion}%
\end{equation}
where r is the relative coordinate of $a$ and $b,$ $u_{lj}^{J}\left(
r\right)  $ is the radial wavefunction\ and $\mathcal{Y}_{JM}^{l}$ is the
spin-angle wavefunction%
\begin{equation}
\mathcal{Y}_{JM}^{l}=\sum_{m\ ,\ M_{a}}\left\langle jmI_{a}M_{a}%
|J{M}\right\rangle \left\vert jm\right\rangle \left\vert I_{a}M_{a}%
\right\rangle ,\ \ \ \ \ \ \mathrm{with}\ \ \ \ \ \left\vert jm\right\rangle
=\sum_{m_{l}\ ,\ M_{b}}Y_{lm_{l}}\left(  \widehat{\mathbf{r}}\right)
\chi_{M_{b}} \label{spinangle}%
\end{equation}
where $\chi_{M_{b}}$\ is the spinor wavefunction of particle $b$ and
$\left\langle jmI_{a}M_{a}|J{M}\right\rangle $ is a Clebsch-Gordan coefficient.

The ground-state wavefunction is normalized so that%
\begin{equation}
\int d^{3}r\ \left\vert \Psi_{JM}\left(  \mathbf{r}\right)  \right\vert ^{2}=%
{\displaystyle\int\limits_{0}^{\infty}}
dr\ \left\vert u_{lj}^{J}\left(  r\right)  \right\vert ^{2}=1. \label{norm}%
\end{equation}

The wavefunctions are calculated using a spin-orbit potential of the form%
\begin{equation}
V(\mathbf{r})=V_{0}(r)+V_{S}(r)\ (\mathbf{l.s})+V_{C}(r)\label{WStot}%
\end{equation}
where $V_{0}(r)$ and $V_{S}(r)$ are the central and spin-orbit interaction,
respectively, and $V_{C}(r)$ is the Coulomb potential of a uniform
distribution of charges:%
\begin{align}
V_{C}(r) &  =\frac{Z_{a}Z_{b}e^{2}}{r}\ \ \ \mathrm{\ for}\ \ \ \ \ r>R_{C}%
\nonumber\\
&  =\frac{Z_{a}Z_{b}e^{2}}{2R_{C}}\left(  3-\frac{r^{2}}{R_{C}^{2}}\right)
\ \ \ \ \mathrm{for}\ \ \ \ r<R_{C},\label{coul_pot}%
\end{align}
where $Z_{i}$ is the charge number of nucleus $i=a,b$.

One can use two kinds of approach to build up the potentials $V_{0}(r)$ and
$V_{S}(r)$. In a Woods-Saxon parametrization they are given by%
\begin{align}
V_{0}(r) &  =V_{0}\ f_{0}(r),\ \ \ \ \ \mathrm{and}\ \ \ \ \ V_{S}%
(r)=-\ V_{S0}\ \left(  \frac{\hbar}{m_{\pi}c}\right)  ^{2}\ \frac{1}{r}%
\ \frac{d}{dr}f_{S}(r)\nonumber\\
\mathrm{with}\ \ \ \ f_{i}(r) &  =\left[  1+\exp\left(  \frac{r-R_{i}}{a_{i}%
}\right)  \right]  ^{-1}\ .\label{centsp}%
\end{align}
The spin-orbit interaction in Eq. \ref{centsp} is written in terms
of the pion Compton wavelength, $\hbar/m_{\pi}c=1.414$ fm. The
parameters $V_{0}$, $V_{S0}$, $R_{0}$, $a_{0},$ $R_{S0}$, and
$a_{S0}$ are adjusted so that the ground state energy $E_B$ (or
the energy of an excited state) is reproduced.

Alternatively, and perhaps more adequate for some situations, one can
construct the potentials using a more microscopic approach. Among these
models, the M3Y interaction is very popular. It has been shown to work quite
reasonably for elastic and inelastic scattering of heavy ions at low and
intermediate energy nuclear collisions \cite{Ber77,Kob84}. It has been applied
to calculations of radiative capture cross sections with relative success
(see, e.g., \cite{Moh98}).

In its simplest form the M3Y interaction is given by two direct terms with
different ranges, and an exchange term represented by a delta interaction:
\begin{equation}
t(s)=A\frac{e^{-\beta_{1}s}}{\beta_{1}s}+B\frac{e^{-\beta_{2}s}}{\beta_{2}%
s}+C\delta(s)\ , \label{tm3y}%
\end{equation}
where one of the possible set for these parameters is given by
\cite{Ber77,Kob84} $A=7999$ MeV, $B=-2134$ MeV, $C=-276$ MeV $fm^{3}$,
$\beta_{1}=4$ $fm^{-1}$, and $\beta_{2}=2.5$ $fm^{-1}$.

The central part of the potential is obtained by a folding of this interaction
with the ground state densities, $\rho_{a}$ and $\rho_{b}$, of the nuclei $a$
and $b$:
\begin{equation}
V_{0}^{M3Y}(r)=\lambda_{0}V^{M3Y}(r)=\lambda_{0}\int d^{3}r_{1}\;d^{3}%
r_{2}\;\rho_{a}(r_{1})\rho_{b}(r_{2})\;t(s)\ ,\label{v0m3y}%
\end{equation}
with $s=\left\vert \mathbf{r}+\mathbf{r}_{2}-\mathbf{r}_{1}\right\vert $.
$\lambda_{0}$ is a normalization factor which is close to unity. We assume
that the densities $\rho_{i}$\ are spherically symmetric. The nuclear
densities can be taken from e.g. Ref. \cite{VJV87} \ for the charge matter
densities. To obtain the matter density one can use the relation $\left\langle
r_{\mathrm{m}}^{2}\right\rangle ^{1/2}=\sqrt{\left\langle r_{\mathrm{ch}}%
^{2}\right\rangle -\left(  0.85\right)  ^{2}}$, where $\left\langle
r_{\mathrm{ch}}^{2}\right\rangle ^{1/2}$ and $\left\langle r_{\mathrm{m}}%
^{2}\right\rangle ^{1/2}$ are the charge and matter rms radii of the nucleus
and the proton radius is taken as 0.85 fm.

The spin-orbit part of the optical potential is parametrized as
\begin{equation}
V_{S}^{M3Y}(r)=-\lambda_{S0}\ \left(  \frac{\hbar}{m_{\pi}c}\right)
^{2}\ \frac{1}{r}\ \frac{d}{dr}V^{M3Y}(r). \label{vsm3y}%
\end{equation}

The bound-state wavefunctions are calculated by solving the radial
Schr\"{o}dinger equation%
\begin{equation}
-\frac{\hbar^{2}}{2m_{ab}}\left[  \frac{d^{2}}{dr^{2}}-\frac{l\left(
l+1\right)  }{r^{2}}\right]  u_{lj}^{J}\left(  r\right)  +\left[  V_{0}\left(
r\right)  +V_{C}\left(  r\right)  +\left\langle \mathbf{s.l}\right\rangle
\ V_{S0}\left(  r\right)  \right]  u_{lj}^{J}\left(  r\right)  =E_{i}%
u_{lj}^{J}\left(  r\right)  \label{bss}%
\end{equation}
where $\left\langle \mathbf{s.l}\right\rangle =\left[
j(j+1)-l(l+1)-s(s+1)\right]  /2$. This equation must satisfy the
boundary conditions $u_{lj}^{J}\left(  r=0\right)
=u_{lj}^{J}\left(  r=\infty\right) =0$ which is only possible for
discrete energies $E$ corresponding to the bound states of the
nuclear + Coulomb potential.

\section{Continuum states}

The continuum wavefunctions are calculated with the potential model as
described above. The parameters are often not the same as the ones used for
the bound states. The continuum states are now identified by the notation
$u_{Elj}^{J}(r)$, where the (continuous) energy $E$ is related to the relative
momentum $k$ of the system $a+b$ by $E=\hbar^{2}k^{2}/2m_{ab}$.

The radial equation to be solved is the same as Eq. \ref{bss}, but with the
boundary conditions at infinity replaced by (see, e.g, Ref. \cite{Jo83})%
\begin{equation}
u_{Elj}^{J}(r\longrightarrow\infty)=i\sqrt{\frac{m_{ab}}{2\pi k\hbar^{2}}%
}\left[  H_{l}^{(-)}\left(  r\right)  -S_{lJ}H_{l}^{(+)}\left(  r\right)
\right]  \ e^{i\sigma_{l}\left(  E\right)  }\label{uE}%
\end{equation}
where $S_{lJ}=\exp\left[  2i\delta_{lJ}\left(  E\right)  \right]  $, with
$\delta_{lJ}\left(  E\right)  $ being the nuclear phase-shift and $\sigma
_{l}\left(  E\right)  $ the Coulomb one, and%
\begin{equation}
H_{l}^{(\pm)}\left(  r\right)  =G_{l}(r)\pm iF_{l}\left(  r\right)
\ .\label{Hl}%
\end{equation}
$F_{l}$ and $G_{l}$ are the regular and irregular Coulomb wavefunctions. If
the particle $b$ is not charged (e.g., a neutron) the Coulomb functions reduce
to the usual spherical Bessel functions, $j_{l}\left(  r\right)  $ and
$n_{l}\left(  r\right)  $.

At a conveniently chosen large distance $r=R$, outside the range of the
nuclear potential, one can define the logarithmic derivative%
\begin{equation}
\alpha_{lJ}=\left(  \frac{du_{Elj}^{J}/dr}{u_{Elj}^{J}}\right)  _{r=R}%
\ .\label{logdev}%
\end{equation}
The phaseshifts $\delta_{lJ}\left(  E\right)  $ are obtained by
matching the logarithmic derivative with the asymptotic value
obtained with the Coulomb
wavefunctions. This procedure yields%
\begin{equation}
S_{lJ}=\frac{G_{l}^{\prime}-iF_{l}^{\prime}-\alpha_{lJ}\ \left(
G_{l} -iF_{l}\right)
}{G_{l}^{\prime}+iF_{l}^{\prime}-\alpha_{lJ}\ \left(
G_{l}+iF_{l}\right)  },\label{match}%
\end{equation}
where the primes mean derivation with respect to the radial coordinate at the
position $R$.

The continuum wavefunctions are normalized so as to satisfy the relation%
\begin{equation}
\left\langle u_{Elj}^{J}|u_{E^{\prime}l^{\prime}j^{\prime}}^{J^{\prime}%
}\right\rangle =\delta\left(  E-E^{\prime}\right)  \delta_{JJ^{\prime}}%
\delta_{jj^{\prime}}\delta_{ll^{\prime}},\label{cont_norm}%
\end{equation}
what means, in practice, that the continuum wavefunctions
$u_{Elj}(r)$ are normalized to
$-\sqrt{2m_{ab}/\pi\hbar^{2}k}~e^{i\delta_{lJ}}\sin
(kr+\delta_{lJ})$ at large $r.$

A resonance in a particular channel $lJ$ is characterized by%
\begin{equation}
\left.  \frac{d^{2}\delta_{lJ}}{dE^{2}}\right\vert _{E_{lJ}^{R}}%
=0,\ \ \ \ \ \ \mathrm{and}\ \ \ \ \ \ \left.  \frac{d\delta_{lJ}}%
{dE}\right\vert _{E_{lJ}^{R}}>0\ . \label{res1}
\end{equation}
The single-particle width of the resonance can be calculated from%
\begin{equation}
\Gamma_{lJ}^{R}=2\left(  \left.
\frac{d\delta_{lJ}}{dE}\right\vert _{E_{lJ}^{R}}\right)  ^{-1}.
\label{res2}
\end{equation}

\section{Multipole matrix elements}

The operators for electric transitions of multipolarity $\lambda\pi$ are given
by (see, e.g. Ref. \cite{BM69})%
\begin{equation}
\mathcal{O}_{E\lambda\mu}=e_{\lambda}\ r^{\lambda}Y_{\lambda\mu}\left(
\widehat{\mathbf{r}}\right)  ,\label{oe}%
\end{equation}
where the effective charge, which takes into account the
displacement of the center-of-mass, is
\begin{equation}
e_{\lambda}=Z_{b}e\left(  -\frac{m_{a}}{m_{c}}\right)  ^{\lambda}%
+Z_{a}e\left(  \frac{m_{b}}{m_{c}}\right)  ^{\lambda}.\label{el}%
\end{equation}

For magnetic dipole transitions%
\begin{equation}
\mathcal{O}_{M1\mu}=\sqrt{\frac{3}{4\pi}}\mu_{N}\ \left[
e_{M}l_{\mu} +\sum_{i=a,b}g_{i}\left(  s_{i}\right)  _{\mu}\right]
,\ \ \ \ \ \ \ \ \ \ \ \ \ \ e_{M}=\left(
\frac{m_{a}^{2}Z_{a}}{m_{c}^{2}
}+\frac{m_{b}^{2}Z_{b}}{m_{c}^{2}}\right)  \ , \label{om1}
\end{equation}
where $l_{\mu}$ and $s_{\mu}$ are the spherical components of order $\mu$
($\mu=-1,0,1$) of the orbital and spin angular momentum ($\mathbf{l=-}%
i\mathbf{r\times\nabla}$, and $\mathbf{s=\sigma}/2$) and $g_{i}$
are the gyromagnetic factors of particles $a$ and $b$. The nuclear
magneton is given by $\mu_{N}=e\hbar/2m_{N}c.$

The matrix element for the transition $J_{0}M_{0}\longrightarrow
JM$, using the
convention of Ref. \cite{BM69}, is given by%
\begin{equation}
\left\langle JM\left\vert \mathcal{O}_{E\lambda\mu}\right\vert
J_{0} M_{0}\right\rangle =\left\langle
J_{0}M_{0}\lambda\mu|JM\right\rangle \ \frac{\left\langle
J\left\Vert \mathcal{O}_{E\lambda}\right\Vert J_{0}\right\rangle
}{\sqrt{2J+1}}.\label{we}
\end{equation}

From the single-particle wavefunctions one can calculate the reduced matrix
elements $\left\langle lj\left\Vert \mathcal{O}_{E\lambda}\right\Vert
l_{0}j_{0}\right\rangle _{J}$. The subscript $J$ is a reminder that the matrix
element depends on the channel spin $J$, because one can use different
potentials in the different channels. The reduced matrix element $\left\langle
J\left\Vert \mathcal{O}_{E\lambda}\right\Vert J_{0}\right\rangle $ can be
obtained from a standard formula of angular momentum algebra, e.g. Eq. (7.17)
of Ref. \cite{Ed60}. One gets%
\begin{equation}
\left\langle J\left\Vert \mathcal{O}_{E\lambda}\right\Vert J_{0}\right\rangle
=\left(  -1\right)  ^{j+I_{a}+J_{0}+\lambda}\ \left[  \left(  2J+1\right)
\left(  2J_{0}+1\right)  \right]  ^{1/2}\ \left\{
\begin{array}
[c]{ccc}%
j & J & I_{a}\\
J_{0} & j_{0} & \lambda
\end{array}
\right\}  \ \left\langle lj\left\Vert \mathcal{O}_{E\lambda}\right\Vert
l_{0}j_{0}\right\rangle _{J}.\label{joj0}%
\end{equation}

To obtain $\left\langle lj\left\Vert \mathcal{O}_{E\lambda}\right\Vert
l_{0}j_{0}\right\rangle _{J}$ one needs the matrix element $\left\langle
lj\left\Vert r^{\lambda}Y_{\lambda}\right\Vert l_{0}j_{0}\right\rangle _{J}$
for the spherical harmonics, e.g. Eq. (A2.23) of Ref. \cite{La80}. For
$l_{0}+l+\lambda=\mathrm{even}$, the result is%
\begin{equation}
\left\langle lj\left\Vert \mathcal{O}_{E\lambda}\right\Vert l_{0}
j_{0}\right\rangle _{J}=\frac{e_{\lambda}}{\sqrt{4\pi}}\ \left(
-1\right) ^{l_{0}+l+j_{0}-j}\
\frac{\hat{\lambda}\hat{j_{0}}}{\hat{\jmath}} \
<j_{0}\tfrac{1}{2}\lambda0|j\tfrac{1}{2}>\ \int_{0}^{\infty}dr\
r^{\lambda }\ u_{lj}^{J}\left(  r\right)  \
u_{l_{0}j_{0}}^{J_{0}}\left(  r\right)
,\label{lol0}%
\end{equation}
where we use here the notation $\widehat{k}=\sqrt{2k+1}$, and $\widetilde
{k}=\sqrt{k(k+1)}$. For $l_{0}+l+\lambda=\mathrm{odd}$, the reduced matrix
element is null.

Eqs. \ref{we} and \ref{joj0} can also be used for the magnetic
dipole excitations. In comparison with the electric dipole
transitions their cross sections are reduced by a factor of
$v^{2}/c^{2}$, where $v$ is the relative velocity of the $a+b$
system. At very low energies, $v\ll c$, and the $M1$ transitions
will be much smaller than the electric transitions. Only in the
case of sharp resonances, the M1 transitions play a role, e.g. for
the $J=1^{+}$ state in $^{8}$B at $E_{R}=630$ keV above the proton
separation threshold \cite{RO73,KPK87}. However, the potential
model apparently is not good in reproducing the M1 transition
amplitudes \cite{Bar88}. We only treat here the case in which the
particle $b$ is a nucleon. For that one needs the reduced matrix
elements $\left\langle lj\left\Vert \widehat{l}\right\Vert
l_{0}j_{0}\right\rangle _{J}$ and $\left\langle lj\left\Vert
\widehat{\sigma }\right\Vert l_{0}j_{0}\right\rangle _{J}$ which
are, e.g., given by Eqs.
(A2.20) and (A2.19) of Ref. \cite{La80}. For $l=l_{0}$ one obtains%
\begin{align}
&  \left\langle lj\left\Vert \mathcal{O}_{M1}\right\Vert
l_{0}j_{0} \right\rangle _{J}=\left(  -1\right)
^{j+I_{a}+J_{0}+1}\ \sqrt{\frac{3}{4\pi }}\
\widehat{J}\widehat{J}_{0}\left\{
\begin{array}
[c]{ccc}%
j & J & I_{a}\\
J_{0} & j_{0} & 1
\end{array}
\right\}  \mu_{N}\nonumber\\
&  \times\left\{  \frac{1}{\widehat{l}_{0}}e_{M}\left[  \frac{2\widetilde
{j}_{0}}{\widehat{l}_{0}}\left(  l_{0}\delta_{j_{0},\ l_{0}+1/2}+\left(
l_{0}+1\right)  \delta_{j_{0},\ l_{0}-1/2}\right)  +\left(  -1\right)
^{l_{0}+1/2-j}\frac{\widehat{j}_{0}}{\sqrt{2}}\delta_{j_{0},\ l_{0}\pm
1/2}\delta_{j,\ l_{0}\mp1/2}\right]  \right.  \nonumber\\
&  +g_{N}\frac{1}{\widehat{l}_{0}^{2}}\left[  \left(  -1\right)
^{l_{0}+1/2-j_{0}}\widetilde{j}_{0}\delta_{j,\ j_{0}}-\left(  -1\right)
^{l_{0}+1/2-j}\frac{\widehat{j}_{0}}{\sqrt{2}}\delta_{j_{0},\ l_{0}\pm
1/2}\delta_{j,\ l_{0}\mp1/2}\right]  \nonumber\\
&  \left.  +g_{a}\left(  -1\right)
^{I_{a}+j_{0}+J+1}\widehat{J}_{0}
\widehat{J}\widehat{I}_{a}\widetilde{I}_{a}\left\{
\begin{array}
[c]{ccc}%
I_{a} & J & j_{0}\\
J_{0} & I_{a} & 1
\end{array}
\right\}  \right\}  \int_{0}^{\infty}dr\ u_{lj}^{J}\left(
r\right) \ u_{l_{0}j_{0}}^{J_{0}}\left(  r\right)  ,\label{ljolj}
\end{align}
The spin g-factor is $g_{N}=5.586$ for the proton and $g_{N}=-3.826$ for the
neutron. The magnetic moment of the core nucleus is given by $\mu_{a}=g_{a}%
\mu_{N}$. If $l\neq l_{0}$ the magnetic dipole matrix element is zero.

\section{The astrophysical S-factor}

The multipole strength, or response functions, for a particular partial wave,
summed over final channel spins, is defined by%
\begin{align}
\frac{dB\left(  \pi\lambda;\ l_{0}j_{0}\longrightarrow klj\right)  }{dk}  &
=\sum_{J}\frac{\left\vert \left\langle kJ\left\Vert \mathcal{O}_{\pi\lambda
}\right\Vert J_{0}\right\rangle \right\vert ^{2}}{2J_{0}+1}\nonumber\\
&  =\sum_{J}\left(  2J+1\right)  \ \left\{
\begin{array}
[c]{ccc}%
j & J & I_{a}\\
J_{0} & j_{0} & \lambda
\end{array}
\right\}  ^{2}\ \left\vert \left\langle klj\left\Vert
\mathcal{O}_{\pi\lambda }\right\Vert l_{0}j_{0}\right\rangle
_{J}\right\vert ^{2}, \label{respf}
\end{align}
where $\pi=E$, or $M$.

If the matrix elements are independent of the channel spin, this sum reduces
to the usual single-particle strength $\left\vert \left\langle klj\left\Vert
\mathcal{O}_{\pi\lambda}\right\Vert l_{0}j_{0}\right\rangle \right\vert
^{2}/\left(  2j_{0}+1\right)  $. For transitions between the bound states the
same formula as above can be used to obtain the reduced transition probability
by replacing the continuum wavefunctions $u_{klj}^{J}\left(  r\right)  $ by
the bound state wavefunction $u_{lj}^{J}\left(  r\right)  $. That is,%
\begin{equation}
B\left(  \pi\lambda;\ l_{0}j_{0}J_{0}\longrightarrow ljJ\right)  =\left(
2J+1\right)  \ \left\{
\begin{array}
[c]{ccc}%
j & J & I_{a}\\
J_{0} & j_{0} & \lambda
\end{array}
\right\}  ^{2}\ \left\vert \left\langle lj\left\Vert
\mathcal{O}_{\pi\lambda }\right\Vert l_{0}j_{0}\right\rangle
\right\vert ^{2}. \label{bval}
\end{equation}

For bound state to continuum transitions the total multipole
strength is obtained by summing over all partial waves,
\begin{equation}
\frac{dB\left(  \pi\lambda\right)  }{dE}=\sum_{lj}\frac{dB\left(
\pi \lambda;\ l_{0}j_{0}\longrightarrow klj\right)  }{dE}\ .
\label{dbde}
\end{equation}

The differential form of the response function in terms of the
momentum $E$ is a result of the normalization of the continuum
waves according to Eq. \ref{uE}.

The photo-absorption cross section for the reaction
$\gamma+c\longrightarrow a+c$ is given in terms of the response
function by \cite{Bla62}
\begin{equation}
\sigma_{\gamma}^{(\lambda)}\left(  E_{\gamma}\right) =\frac{\left(
2\pi\right)  ^{3}\left(  \lambda+1\right) }{\lambda\left[  \left(
2\lambda+1\right)  !!\right]  ^{2}}\left(
\frac{m_{ab}}{\hbar^{2}k}\right) \left(  \frac{E_{\gamma}}{\hbar
c}\right)  ^{2\lambda-1}\frac{dB\left( \pi\lambda\right)  }{dE},
\label{sig_lambda}
\end{equation}
where $E_{\gamma}=E+|E_B|$, with $|E_B|$ being the binding energy
of the $a+b$ system.
For transitions between bound states, one has%
\begin{equation}
\sigma_{\gamma}^{(\pi\lambda)}\left(  E_{\gamma}\right)
=\frac{\left( 2\pi\right)  ^{3}\left(  \lambda+1\right)
}{\lambda\left[  \left( 2\lambda+1\right)  !!\right]  ^{2}}\left(
\frac{E_{\gamma}}{\hbar c}\right) ^{2\lambda-1}B\left(
\pi\lambda;\ l_{0}j_{0}J_{0}\longrightarrow ljJ\right)
\delta\left(  E_{f}-E_{i}-E_{\gamma}\right)  , \label{photo}
\end{equation}
where $E_{i}$ ($E_{f}$) is the energy of the initial (final) state.

The cross section for the radiative capture process $a+b\longrightarrow
c+\gamma$ can be obtained by detailed balance \cite{Bla62}, and one gets%
\begin{equation}
\sigma_{(\pi\lambda)}^{(\mathrm{rc})}\left(  E\right)  =\left(
\frac {E_{\gamma}}{\hbar c}\right)  ^{2\lambda-1}\frac{2\left(
2I_{c}+1\right)
}{(2I_{a}+1)(2I_{b}+1)}\sigma_{\gamma}^{(\lambda)}\left(
E_{\gamma}\right)  . \label{radcap}
\end{equation}
The total capture cross section $\sigma_{\mathrm{nr}}$ is
determined by the capture to all bound states with the single
particle spectroscopic factors $C^{2}S_{i}$ in the final nucleus
\begin{equation}
\sigma_{\mathrm{nr}}\left(  E\right)  =\sum_{i,\pi,\lambda}(C^{2}
S)_{i}\ \sigma_{(\pi\lambda),i}^{(\mathrm{rc})}\left(  E\right)  .
\label{NR}
\end{equation}
Experimental information or detailed shell model calculations have to be
performed to obtain the spectroscopic factors $(C^{2}S)_{i}$. For example, the
code OXBASH \cite{bro84} can used for this purpose.

For charged particles the astrophysical S-factor for the direct
capture from a continuum state to the bound state is defined as
\begin{equation}
S^{(c)}\left(  E\right)  =E\ \sigma_{\mathrm{nr}}\left(  E\right)
\;\exp\left[ 2\pi\eta\left(  E\right)  \right]  ,\ \ \ \ \ \ \ \ \
\ \ \ \ \mathrm{with} \ \ \ \ \ \ \ \eta\left(  E\right)
=Z_{a}Z_{b}e^{2}/\hbar v, \label{s_lambda}
\end{equation}
where $v$ is the relative velocity between $a$ and $b$.

\section{Nuclear reaction rates in stellar environments}

The nuclear reaction rate, measuring the number of reactions per
particle pair, $a+b$, per second in the stellar environment can be
calculated from the nuclear cross section $\sigma$ for a given
reaction by folding it with the velocity distribution of the
particles involved. In most astrophysical applications the nuclei
are in a thermalized plasma, yielding a Maxwell-Boltzmann velocity
distribution. The astrophysical reaction rate $R$ at a temperature
$T$ can then be written as~\cite{Fo67}
\begin{equation}
R(T)=\frac{n_{a}n_{b}}{1+\delta_{ab}}\left\langle \sigma v\right\rangle \ ,
\end{equation}
where $n_{i}$ is the number density of the nuclear species $i$. The
denominator takes care of the special case of two identical nuclei in the
entrance channel. The quantity $\left\langle \sigma v\right\rangle $ is given
by%
\begin{equation}
\left\langle \sigma v\right\rangle =\left(  \frac{8}{\pi
m_{ab}}\right)
^{1/2}\frac{1}{(k_{\mathrm{B}}T)^{3/2}}\int_{0}^{\infty}\sigma(E)\
E\ \exp \left(  -\frac{E}{k_{\mathrm{B}}T}\right)  \,dE\
,\label{ver}
\end{equation}
with $k_{\mathrm{B}}$ the Boltzmann constant.

The threshold behavior of radiative capture cross sections is fundamental in
nuclear astrophysics because of the small projectile energies in the
thermonuclear region. For example, for neutron capture near the threshold the
cross section can be written \cite{Bla62} as%
\begin{equation}
\sigma_{if}=\frac{\pi}{k^{2}}\frac{-4kR\
\mathrm{Im}\alpha_{0}}{\left\vert \alpha_{0}\right\vert ^{2}}\
,\label{CST}
\end{equation}
where $\alpha_{0}$ is the logarithmic derivative for the $s$ wave, given by
Eq. \ref{logdev}. Since $\alpha_{0}$ is only weakly dependent on the
projectile energy, one obtains for low energies the well--known $1/v$--behavior.

With increasing neutron energy higher partial waves with $l>0$ contribute more
significantly to the radiative capture cross section. Thus the product $\sigma
v$ becomes a slowly varying function of the neutron velocity and one can
expand this quantity in terms of $v$ or $\sqrt{E}$ around zero energy:
\begin{equation}
\sigma v=S^{(n)}(0)+\dot{S}^{(n)}(0)\sqrt{E}+{1\over 2} \ddot{S}^{(n)}(0)E+\ldots\ .\label{AS}%
\end{equation}
The quantity $S^{(n)}(E)=\sigma v$ is the astrophysical S--factor
for neutron--induced reactions and the dotted quantities represent
derivatives with respect to $E^{1/2}$, i.e.,
$\dot{S}^{(n)}=2\sqrt{E} {\ dS^{(n)}\over dE}$ and
$\ddot{S}^{(n)}=4E\ {d^2S^{(n)}\over dE^2} + 2\ {dS^{(n)}\over
dE}$. Notice that the above astrophysical S--factor for
neutron--induced reactions is different from that for
charged--particle induced reactions. In the astrophysical
S--factor for charged--particle induced reactions also the
penetration factor through the Coulomb barrier has to be
considered (Eq. \ref{s_lambda}).

Inserting this into Eq.~\ref{ver} we obtain for the reaction rate for
neutron--induced reactions%
\begin{equation}
\left\langle \sigma v\right\rangle =S(0)+\left(
\frac{4}{\pi}\right)
^{\frac{1}{2}}\dot{S}(0)(k_{\mathrm{B}}T)^{\frac{1}{2}}+\frac{3}{4}\ddot
{S}(0)k_{\mathrm{B}}T+\ldots.\label{RRN}
\end{equation}

In most astrophysical neutron--induced reactions, neutron s--waves will
dominate, resulting in a cross section showing a 1/$v$--behavior (i.e.,
$\sigma(E)\propto1/\sqrt{E}$). In this case, the reaction rate will become
independent of temperature, $R=\mathrm{const}$. Therefore it will suffice to
measure the cross section at one temperature in order to calculate the rates
for a wider range of temperatures. The rate can then be computed very easily
by using%
\begin{equation}
R=\left\langle \sigma v\right\rangle =\left\langle \sigma\right\rangle
_{T}v_{T}=\mathrm{const}\ ,
\end{equation}
with%
\begin{equation}
v_{T}=\left(  \frac{2kT}{m}\right)  ^{1/2}\ .
\end{equation}

The mean lifetime $\tau_{\mathrm{n}}$ of a nucleus against neutron capture,
i.e., the mean time between subsequent neutron captures is inversely
proportional to the available number of neutrons $n_{\mathrm{n}}$ and the
reaction rate $R_{\mathrm{n\gamma}}$:
\begin{equation}
\tau_{\mathrm{n}}=\frac{1}{n_{\mathrm{n}}R_{\mathrm{n\gamma}}}\ .
\end{equation}
If this time is shorter than the beta--decay half--life of the nucleus, it
will be likely to capture a neutron before decaying (r-process). In this
manner, more and more neutrons can be captured to build up nuclei along an
isotopic chain until the beta--decay half--life of an isotope finally becomes
shorter than $\tau_{\mathrm{n}}$. With the very high neutron densities
encountered in several astrophysical scenarios, isotopes very far--off
stability can be synthesized.

For low $|E_B|$-values, e.g. for halo-nuclei, the simple $1/v$-law
does not apply anymore. A significant deviation can be observed if
the neutron energy is of the order of the $|E_B|$-value. In this
case the response function in Eq. \ref{dbde} can be calculated
analytically under simplifying assumptions (see Ref.\cite{BS92}).
For direct capture to weakly bound final states, the bound--state
wave function $u_{lj}(r)$ decreases very slowly in the nuclear
exterior, so that the contributions come predominantly from far
outside the nuclear region, i.e., from the \textit{nuclear
halo\/}. For this asymptotic region the scattering and bound wave
functions in Eq.~\ref{psi_expansion} can be approximated by their
asymptotic expressions neglecting the nuclear
potential \cite{ots94}%
\begin{equation}
u_{l}(kr)\propto j_{l}(kr),\ \ \ \ \ \ \ \ \ u_{l_{0}}(r)\propto
h_{l_{0} }^{(+)}(i\eta r)\ ,\nonumber
\end{equation}
where $j_{l}$ and $h_{l_{0}}^{(+)}$ are the spherical Bessel, and
the Hankel function of the first kind, respectively. The
separation energy $|E_B|$ in the exit channel is related to the
parameter $\eta$ by $|E_B|=\hbar^{2}\eta ^{2}/(2m_{ab})$.

Performing the calculations of the radial integrals in Eq. \ref{lol0}, one
readily obtains the energy dependence of the radiative capture cross section
for halo nuclei \cite{BS92,ots94}. For example, for a transition
s$\longrightarrow$p it becomes%
\begin{equation}
\sigma_{(E1)}^{(\mathrm{rc})}(\mathrm{s}\rightarrow\mathrm{p})\propto\frac
{1}{\sqrt{E}}\frac{(E+3|E_B|)^{2}}{E+|E_B|}\ , \label{qqq1}
\end{equation}
while a transition p$\rightarrow$s has the energy dependence
\begin{equation}
\sigma_{(E1)}^{(\mathrm{rc})}(\mathrm{p}\rightarrow\mathrm{s})\propto
\frac{\sqrt{E}}{E+|E_B|}\ .\label{qqq2}
\end{equation}
If $E\ll |E_B|$ the conventional energy dependence is recovered.
From the above equations one obtains that the reaction rate is not
constant (for s-wave capture) or proportional to $T$ (for p-wave
capture) in the case of small $|E_B|$-values. These general
analytical results can be used as a guide for interpreting the
numerical calculations involving neutron halo nuclei.

In the case of charged particles $S(E)$ is expected to be a slowly varying
function in energy for non-resonant nuclear reactions. In this case, $S\left(
E\right)  $ can be expanded in a McLaurin series,%
\begin{equation}
S\left(  E\right)  =S\left(  0\right)  +\dot{S}\left(  0\right)  E+\frac{1}%
{2}\ddot{S}\left(  0\right)  E^{2}+\cdots\label{sexp}%
\end{equation}

Using this expansion in Eq. \ref{ver} and approximating the
product of the exponentials $\exp\left(  -E/k_{\mathrm{B}}T\right)
$ and $\exp\left[ 2\pi\eta\left(  E\right)  \right]  $ by a
Gaussian centered at the energy $E_{0}$, Eq. \ref{ver} \ can be
evaluated as \cite{Rol88}
\begin{equation}
\left\langle \sigma v\right\rangle =\left(  \frac{2}{m_{ab}}\right)
^{1/2}\text{ }\frac{\Delta}{\left(  kT\right)  ^{3/2}}\ S_{\mathrm{eff}%
}\left(  E_{0}\right)  \exp\left(  -\frac{3E_{0}}{kT}\right)
\label{svap}
\end{equation}
with%
\begin{equation}
S_{\mathrm{eff}}\left(  E_{0}\right)  =S\left(  0\right)  \left[
1+\frac {5}{12\tau}+\frac{\dot{S}\left(  0\right)  }{S\left(
0\right)  }\left( E_{0}+\frac{35E_{0}}{12\tau}\right)
+\frac{\ddot{S}\left(  0\right) }{2S\left(  0\right)  }\left(
E_{0}^{2}+\frac{89E_{0}^{2}}{12\tau}\right)
\right]  \ .\label{svap2}%
\end{equation}

The quantity $E_{0}$\ defines the effective mean energy for thermonuclear
fusion reactions at a given temperature $T$,%
\begin{equation}
E_{0}=1.22\left(  Z_{a}^{2}Z_{b}^{2}m_{ab}T_{6}^{2}\right)  ^{1/2}%
\ \ \mathrm{keV}\ , \label{e0}%
\end{equation}
where $T_{6}$ measures the temperature in 10$^{6}$ K. The quantities $\tau$
and $\Delta$ are given by%
\begin{equation}
\tau=\frac{3E_{0}}{kT},\ \ \ \ \ \ \ \Delta=\frac{4}{\sqrt{3}}\left(
E_{0}kT\right)  ^{1/2}\ . \label{tD}%
\end{equation}

An analytical insight of the cross sections and astrophysical S-factors for
proton-halo nuclei can also be developed (see, e.g., Ref. \cite{For03}).
However, due to the Coulomb field the expressions become more complicated. The
analytical formulas for direct capture cross sections involving neutron and
proton halo nuclei are very useful to interpret the results obtained in a
numerical calculation.

For the case of resonances, where $E_{r}$ is the resonance energy,
we can approximate $\sigma\left(  E\right)  $ by a Breit-Wigner
resonance formula
\cite{Bre36,Bla62}:%
\begin{equation}
\sigma_{\mathrm{r}}(E)=\frac{\pi\hbar^{2}}{2\mu E}\frac{\left(  2J_{R}%
+1\right)  }{(2J_{a}+1)(2J_{b}+1)}\frac{\Gamma_{\mathrm{p}}\Gamma
_{\mathrm{\gamma}}}{\left(  E_{\mathrm{r}}-E\right)  ^{2}+\left(
\Gamma_{\mathrm{tot}}/2\right)  ^{2}}\ ,\label{BW}%
\end{equation}
where $J_{R}$, $J_{a}$, and $J_{b}$ are the spins of the resonance and the
nuclei $a$ and $b$, respectively, and the total width $\Gamma_{\mathrm{tot}}$
is the sum of the particle decay partial width $\Gamma_{\mathrm{p}}$ and the
$\gamma$-ray partial width $\Gamma_{\gamma}$. The particle partial width, or
entrance channel width, $\Gamma_{\mathrm{p}}$ can be expressed in terms of the
single-particle spectroscopic factor $S$ and the single-particle width
$\Gamma_{\mathrm{s.p.}}$ of the resonance state~\cite{sch63}
\begin{equation}
\Gamma_{\mathrm{p}}=C^{2}S\times\Gamma_{\mathrm{s.p.}}\ ,
\end{equation}
where $C$ is the isospin Clebsch-Gordan coefficient. The single-particle width
$\Gamma_{\mathrm{s.p.}}$ can be calculated from the scattering phase shifts of
a scattering potential with the potential parameters being determined by
matching the resonance energy (see Eq. \ref{res2}).

The gamma partial widths $\Gamma_{\mathrm{\gamma}}$ are calculated from the
electromagnetic reduced transition probabilities B($J_{i}\rightarrow J_{f}$;L)
which carry the nuclear structure information of the resonance states and the
final bound states \cite{bru77}. The reduced transition rates can be computed
within the framework of the shell model.

Most of the typical transitions are M1 or E2 transitions. For these the
relations are
\begin{equation}
\Gamma_{\mathrm{E2}}[\mathrm{{eV}}]=8.13\times10^{-7}\ E_{\gamma}%
^{5}\mathrm{\ [{MeV}]\ \ }B(E2\mathrm{)\ [{e^{2}fm^{4}}]} \label{gl1}%
\end{equation}
and
\begin{equation}
\Gamma_{\mathrm{M1}}[\mathrm{{eV}}]=1.16\times10^{-2}\ E_{\gamma}%
^{3}\mathrm{\ [{MeV}]\ }B(M1)\mathrm{\ }[\mu_{N}^{2}]\mathrm{\ .} \label{gl2}%
\end{equation}

For the case of narrow resonances, with width $\Gamma\ll E_{r}$,
the Maxwellian exponent $\exp\left(  -E/k_{\mathrm{B}}T\right)  $
can be taken out
of the integral, and one finds%
\begin{equation}
\left\langle \sigma v\right\rangle =\left(  \frac{2\pi}{m_{ab}kT}\right)
^{3/2}\hbar^{2}\left(  \omega\gamma\right)  _{R}\exp\left(  -\frac{E_{r}}%
{kT}\right)  \ , \label{svres}%
\end{equation}
where the resonance strength is defined by%
\begin{equation}
\left(  \omega\gamma\right)  _{R}=\frac{2J_{R}+1}{(2J_{a}+1)(2J_{b}%
+1)}\ \left(  1+\delta_{ab}\right)  \ \frac{\Gamma_{\mathrm{p}}\ \Gamma
_{\gamma}}{\Gamma_{\mathrm{tot}}}. \label{svres2}%
\end{equation}

For broad resonances Eq. \ref{ver} is usually calculated numerically. An
interference term has to be added. The total capture cross section is then
given by~\cite{rol74}
\begin{equation}
\sigma(E)=\sigma_{\mathrm{nr}}(E)+\sigma_{\mathrm{r}}(E)+2\left[
\sigma_{\mathrm{nr}}(E)\sigma_{\mathrm{r}}(E)\right]  ^{1/2}\mathrm{cos}%
[\delta_{\mathrm{R}}(E)]\ .\label{eq-int}%
\end{equation}
In this equation $\delta_{\mathrm{R}}(E)$ is the resonance phase shift. Close
to a resonance, the phase shift approaches the value $\pi/2.$ Thus, close to a
resonance one can use the expansion%
\[
\delta_{\mathrm{R}}(E)\simeq\frac{\pi}{2}-\left(  E_{r}-E\right)
\left. \frac{d\delta}{dE}\right\vert _{R}\ .
\]
Thus, using the definition given by Eq. \ref{res2}, one has%
\begin{equation}
\delta_{\mathrm{R}}(E)=\mathrm{arctan}~\frac{\Gamma}{2(E-E_{\mathrm{R}})}\ .
\end{equation}

Only the contributions with the same angular momentum of the incoming wave
interfere in Eq.~\ref{eq-int}.

\section{Computer program and user's manual}

All nuclear quantities, either known from experiments or
calculated from a model, as well as the conditions realized in the
experiment, are explicitly specified as input parameters. The
program RADCAP then computes the potentials, bound state energies,
phase-shifts, transition probabilities, photo-dissociation cross
sections and astrophysical $S$-factors.

The units used in the program are fm (femtometer) for distances and MeV for
energies. The output cross sections are given in millibarns and the S-factors
in eV.b.

The program is very fast and does not require a complicated input. It asks the
user the calculation one wants to perform. It is divided in 5 modules and one
enters the following options when prompted on the screen:

1 - for the calculation of M3Y potential,

2 - for the calculation of energy and wavefunction of bound states,

3 - for the calculation of reduced transition probabilities between bound states,

4 - for the calculation of phase shifts and wavefunctions of continuum states,

5 - for the calculation of astrophysical S-factors, response functions,
photo-dissociation, and direct capture cross sections.

For each option, a different subroutine is used: \ 1=OMP\_M3Y, 2=EIGEN,
3=BVALUE, 4=CONT and 5=DICAP.

The inputs can be commented by using the symbol "*" at the first
position of an input line.

Note that the angular momenta described in the text have the
following correspondence in the program: $l,\ l_{0}$ (L, L0); $j,\
j_{0}$ (J, J0), $I_{a},\ I_{b}\left(  s\right)  $ (AIA, AIB), $J,\
J_{0}$ (AICF,$\ $AIC). The program notation for the other
variables are easy to recognize (see test cases below).

\subsection{The M3Y potential}

To obtain the M3Y potential one selects the option 1. This calls
the subroutine OMP\_M3Y. The input file is named M3Y.INP. If \ the
densities are not parametrized either by a Gaussian or by a
Woods-Saxon function, one can enter them in the input file
DENS.INP in rows of $r\times\rho_{a}(r)\times \rho_{b}(r)$. The
first line of this input file should contain only the number of
points in $r$. The function DENS\_READ will read those densities
and interpolate them for use in the other routines. An example is
calculation of the M3Y potential for the system
$\mathrm{p}+^{7}\mathrm{Be}$ as follows. For the proton one can
use a Gaussian density with radius parameter $R=0.7$ fm. For
$^{7}\mathrm{Be}$ a Gaussian density parametrization can be used
\cite{VJV87} with radius parameter $R=1.96.$ An appropriate input
file is listed below.

{\small **********************************************}%

{\small * ******** Input of subroutine OMP\_M3Y *******}

{\small * IOPT = Option for densities: = 0 Gaussian or
Woods-Saxon,}

{\small *  \ \hspace{0.09in} \ \hspace{0.09in} \ \hspace{0.09in}=
1 densities entered in DENS.INP}

{\small * NPNTS = number of points in the radial mesh ($<$ 10000)}

{\small * RMAX = maximum radius size (fm)  ( $<$ 250 fm)}

{\small * IOPT \ \hspace{0.09in}NPTS \ \hspace{0.09in}RMAX}

{\small \ \hspace{0.09in}0 \ \hspace{0.09in}100 \ \hspace{0.09in}10.}

{\small * If IOPT = 0, enter density parameters:}

{\small * R1, D1 = Woods-Saxon form (radius and diffuseness)}

{\small * R2, D2 = Same but for density of nucleus 2}

{\small * For Gaussian densities, enter D1=0, or D2=0}

{\small * R1 \ \hspace{0.09in} D1 \ \hspace{0.09in}R2 \
\hspace{0.09in} D2}

{\small \ \hspace{0.09in}0.7 \ \hspace{0.09in} 0. \ \hspace{0.09in} 1.96
\ \hspace{0.09in}0. }

{\small * Mass numbers }

{\small * A1 \ \hspace{0.09in}A2}

{\small \ \hspace{0.09in}1. \ \hspace{0.09in}7. }

{\small **********************************************}%

The subroutine OMP\_M3Y builds up the nuclear potential and calls
the subroutine TWOFOLD which does all the work. It does the
integration appearing in Eq. \ref{v0m3y}. The outputs will appear
in files OMP.TXT and OMP.INP. The later one is for use as input of
the M3Y potential by the other subroutines (if required). Figure\
1 shows a plot of the potential obtained with this input.

\begin{figure}
[ptb]
\begin{center}
\includegraphics[
height=2.7968in,
width=3.4169in
]%
{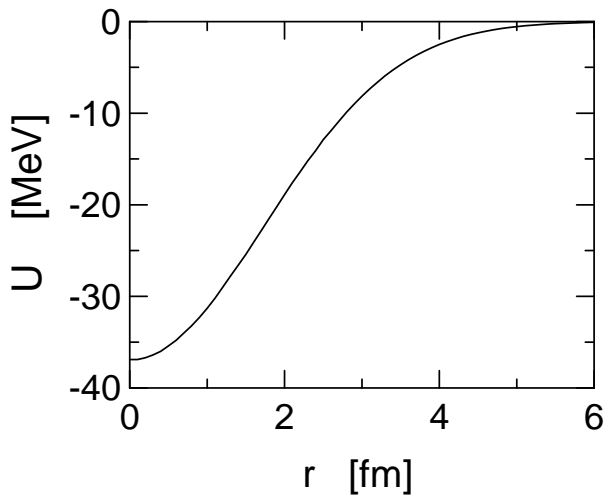}%
\caption{M3Y potential for the system $p+^{7}\mathrm{Be}$.}%
\label{f1}%
\end{center}
\end{figure}

\subsection{Eigenfunctions and energies}

The option 2 calls the subroutine EIGEN. If the real part of the potential is
given as an input file OMP.INP (e.g. the one generated by the subroutine
OMP\_M3Y) it should be written in rows of $r\times V(r)$. The first line of
this input file should contain only the number of points in $r$. The function
OMP\_READ will read and interpolate the potential for use in the other
routines. The subroutine DERIVATIVE calculates its derivative to be used in
the calculation of Eq. \ref{vsm3y}. Let us assume we want to find the ground
state of $^{8}$B. The 2$^{+}$ ground state of $^{8}$B\ can be described as a
p3/2 proton coupled to the 3/2$^{-}$ ground state $^{7}$Be. The subroutine
POTENT builds up the potential. An example of the input file, named EIGEN.INP,
which uses a Woods-Saxon potential, is shown as follows.

{\small **********************************************}%

{\small * ******** Input of subroutine EIGEN *******}

{\small * IOPT = option for potentials: 1 (2) for Woods-Saxon
(M3Y)}

{\small * NPNTS = no. of integration points in radial coordinate (
$<$ 10000)}

{\small * RMAX = maximum radius size ( $<$ 250 fm)}

{\small * IOPT \ \hspace{0.09in} NPTS \ \hspace{0.09in}RMAX}

{\small \ \hspace{0.09in}1 \ \hspace{0.09in}9999 \ \hspace{0.09in} 250.}

{\small * N\_0 = nodes of the Wave Function (exclude origin) }

{\small * J0  = single-particle angular momentum }

{\small * L0  = orbital angular momentum}

{\small * N\_0 \ \hspace{0.09in} J0 \ \hspace{0.09in}L0}

{\small \ \hspace{0.09in}0 \ \hspace{0.09in}1.5 \ \hspace{0.09in} 1}

{\small * If IOPT = 1, enter:}

{\small * V0 = depth of central  potential}

{\small * VS0 = depth of spin-orbit potential }

{\small * R0 = radius of the potential }

{\small * AA = diffuseness of the potential  }

{\small * RS0 = radius of the spin-orbit potential }

{\small * AAS = diffuseness of the spin-orbit potential  }

{\small * RC = Coulomb radius (usually, RC = R0) }

{\small
*-----------------------------------------------------------------}

{\small * WS = V\_0 f(r,R0,AA) - V\_S0 (l.s) (r\_0\symbol{94}2/r)
d/dr f(r,RS0,AAS) }

{\small * f(r,R0,a) = [ 1 + exp((r-R\_0)/a) ]\symbol{94}(-1)}

{\small * r\_0 = 1.4138 fm is the Compton wavelength of the pion.
}

{\small
*-----------------------------------------------------------------}

{\small * V0 \ \hspace{0.09in} R0 \ \hspace{0.09in} AA \
\hspace{0.09in} VS0 \ \hspace{0.09in} RS0 \ \hspace{0.09in} AAS \
\hspace{0.09in} RC}

{\small \ \hspace{0.09in}-44.658 \ \hspace{0.09in}2.391 \ \hspace{0.09in}0.52
\ \hspace{0.09in}-9.8 \ \hspace{0.09in} 2.391 \ \hspace{0.09in} 0.52
\ \hspace{0.09in} 2.391 }

{\small * If IOPT = 2, or else (but not 1), enter FC, FSO and RC}

{\small * (in this case, insert a '*' sign in above row, or delete
it)}

{\small * FC = multiplicative factor of central part of M3Y
potential}

{\small * FSO = multiplicative factor of spin-orbit part of M3Y
potential}

{\small * RC = Coulomb radius }

{\small * FC \ \hspace{0.09in} FSO \ \hspace{0.09in} RC}

{\small * 1.5 \ \hspace{0.09in}0.2 \ \hspace{0.09in} 2.391}

{\small * Z1, Z2 = charges of the nuclei}

{\small * A1, A2 = masses of the nuclei (in nucleon mass units) }

{\small * Z1 \ \hspace{0.09in}A1 \ \hspace{0.09in}Z2 \
\hspace{0.09in}A2}

{\small \ \hspace{0.09in}1. \ \hspace{0.09in}1. \ \hspace{0.09in}4.
\ \hspace{0.09in}7. }

{\small **********************************************}%

In the example input shown above the potential parameters were chosen so as to
reproduce the proton separation energy in $^{8}\mathrm{B}$ which is equal to
0.136 MeV. If the M3Y potential was used, an input of the parameters
$\lambda_{0}$ (FC), $\lambda_{SO}$ (FSO), and $R_{C}$ in Eqs. \ref{coul_pot},
\ref{v0m3y} and \ref{vsm3y} is needed. Note that this input line was
commented, as we did not use it.

The calculations are mainly done in the subroutine BOUNDWAVE which solves the
Schr\"{o}dinger equation for the bound-state problem. When Woods-Saxon
potentials are used they are constructed in the routine POTENTIAL.

The output of the wavefunction will be printed in EIGEN.TXT and GSWF.INP. The
later is prepared for use as input wavefunction for the subroutine BVALUE
(reduced transition probabilities), or the subroutine DICAP (direct capture
subroutine). The solid line in Figure\  2 shows the ground state wavefunction
of $^{8}\mathrm{B}$ obtained with this input.

\begin{figure}
[ptb]
\begin{center}
\includegraphics[
height=2.7985in,
width=3.4731in
]
{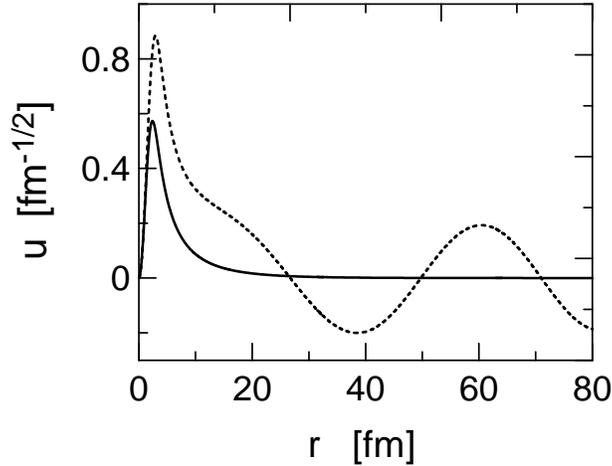}
\caption{The solid line shows the ground state wavefunction of $^{8}%
\mathrm{B}$ in the potential model and the dotted line shows the real part of
the wavefunction for the 1$^{+}$ resonance in $^{8}$B at 630 keV (see text for
details).}%
\label{f2}%
\end{center}
\end{figure}

\subsection{Reduced transition probabilities}

The option 3 calls the subroutine BVALUE which calculates reduced
transition probabilities. To make and example with
$^{8}\mathrm{B}$ we artificially generate a p3/2, 1$^{+}$ state,
with excitation energy of 90 keV. This can be obtained by changing
the WS potential input of EIGEN.INP to the values shown below

{\small * V0 \ \hspace{0.09in} R0 \ \hspace{0.09in} AA \
\hspace{0.09in} VS0 \ \hspace{0.09in} RS0 \ \hspace{0.09in} AAS \
\hspace{0.09in} RC}

{\small \ \hspace{0.09in}-30.55 \ \hspace{0.09in} 2.95 \ \hspace{0.09in} 0.52
\ \hspace{0.09in} -8.53 \ \hspace{0.09in} 2.95 \ \hspace{0.09in} 0.52
\ \hspace{0.09in} 2.95 }

The output yields a state with energy of $-0.05$ MeV.

The input file with the option 3 is to be given in BVALUE.INP. To calculate
the reduced transition \ probability the input file should look like the
example below.

{\small **********************************************}%

{\small * ******** Input of subroutine BVALUE *******}

{\small * AIA = spin of the particle A (core)}

{\small * AIB = intrinsic spin of the particle B }

{\small * AIC = total angular momentum of the ground state of C =
A + B}

{\small * (channel spin)}

{\small * J0 = single particle angular momentum of B respective to
A}

{\small * L0 = relative orbital angular momentum of the ground
state}

{\small * AIA \ \hspace{0.09in}AIB \ \hspace{0.09in}AIC \
\hspace{0.09in}J0 \ \hspace{0.09in}L0}

{\small \ \hspace{0.09in}1.5 \ \hspace{0.09in} 0.5 \ \hspace{0.09in} 2
\ \hspace{0.09in}1..5 \ \hspace{0.09in} 1 }

{\small * N\_1 = nodes of the excited state wave function (exclude
origin) }

{\small * J = single-particle angular momentum }

{\small * L = orbital angular momentum}

{\small * AICF = spin of the excited state after all angular
momentum coupling}

{\small * (channel spin)}

{\small * N\_1 \ \hspace{0.09in}J \ \hspace{0.09in}L \
\hspace{0.09in}AICF }

{\small \ \hspace{0.09in}0 \ \hspace{0.09in}1.5 \ \hspace{0.09in}1
\ \hspace{0.09in}1}

{\small * JOPT = 1 (0) if final state angular momentum, AICF, is
(is not) to be}

{\small * summed over all possible values. If JOPT=1, AICF in the}

{\small * previous line can be entered as any value.}

{\small * JOPT}

{\small \ \hspace{0.09in}0 }

{\small * Z1, Z2 = charges of the nuclei}

{\small * A1, A2 = masses of the nuclei (in nucleon mass units) }

{\small * Z1 \ \hspace{0.09in}A1 \ \hspace{0.09in}Z2 \
\hspace{0.09in}A2}

{\small \ \hspace{0.09in}1. \ \hspace{0.09in}1. \ \hspace{0.09in} 4.
\ \hspace{0.09in}7. }

{\small * IOPT = option for potentials: 1 (2) for Woods-Saxon
(M3Y)}

{\small * Integration parameters for radial wavefunctions:}

{\small * NPNTS = no. of integration points in radial coordinate (
$<$ 10000)}

{\small * RMAX = maximum radius size ( 250 fm)}

{\small * IOPT NPTS RMAX}

{\small \ \hspace{0.09in}1 \ \hspace{0.09in}9999 \ \hspace{0.09in}250.}

{\small * V0 = depth of central potential }

{\small * R0 = radius of the central potential }

{\small * AA = diffuseness of the central potential }

{\small * VS0 = depth of spin-orbit potential }

{\small * RS0 = radius of the spin-orbit potential }

{\small * AAS = diffuseness of the spin-orbit potential }

{\small * RC = Coulomb radius (usually, RC = R0) }

{\small
*-----------------------------------------------------------------}

{\small * WS = V\_0 f(r,R0,AA) - V\_S0 (l.s) (r\_0\symbol{94}2/r)
d/dr f(r,RS0,AAS) }

{\small * f(r,R0,a) = [ 1 + exp((r-R\_0)/a) ]\symbol{94}(-1)}

{\small * r\_0 = 1.4138 fm is the Compton wavelength of the pion.
}

{\small
*-----------------------------------------------------------------}

{\small * V0 \ \hspace{0.09in}R0 \ \hspace{0.09in}AA \
\hspace{0.09in}VS0 \ \hspace{0.09in}RS0 \ \hspace{0.09in}AAS \
\hspace{0.09in}RC}

{\small \ \hspace{0.09in}-30.55 \ \hspace{0.09in}2.95 \ \hspace{0.09in}0.52
\ \hspace{0.09in}-8.53 \ \hspace{0.09in}2.95 \ \hspace{0.09in}0.52
\ \hspace{0.09in}2.95 }

{\small * If IOPT = 2, or else (but not 1), enter FC, FSO and RC:}

{\small * (in this case, insert a '*' sign in above row, or delete
it)}

{\small * FC = multiplicative factor of central part of M3Y
potential}

{\small * FSO = multiplicative factor of spin-orbit part of M3Y
potential}

{\small * RC = Coulomb radius }

{\small * FC \ \hspace{0.09in}FSO \ \hspace{0.09in}RC}

{\small * 1.5 \ \hspace{0.09in}0.2 \ \hspace{0.09in}2.391}

{\small * MP = multipolarity: 0 (M1), 1 (E1), 2 (E2) }

{\small * MP}

{\small \ \hspace{0.09in}2}

{\small * GA = magnetic moment (in units of the nuclear magneton)
of }

{\small * particle A (core)}

{\small * GB = magnetic moment of particle B}

{\small * GA \ \hspace{0.09in}GB}

{\small \ \hspace{0.09in}2.79 \ \hspace{0.09in}-1.7}

{\small **********************************************}%

The output of this run yields $B\left(  E2;\ i\longrightarrow f\right)  =3.76$
e$^{2}$ fm$^{4}$. The spectroscopic factors for the initial and final states
are taken as the unity. If they are known one just multiply this result by
their corresponding values.

The bound state is calculated by the routine BOUNDWAVE and the 3-j and 6-j
coefficients are calculated in the routines THREEJ and SIXJ, respectively.

\subsection{Phase-shifts and resonances}

If one uses the option 4 the program will calculate the scattering
phase-shifts for a given set of potential parameters and angular momentum
quantum numbers for the continuum waves. For example, one might want to
calculate the phase-shifts for the p$+^{8}$Be\ system in the energy interval
$E=0-3$ MeV. The input file CONT.INP could be written as follows.

{\small **********************************************}%

{\small * ******** \ \hspace{0.09in} Input of subroutine CONT \
\hspace {0.09in} *******}

{\small * IOPT = option for potentials: 1 (2) for Woods-Saxon
(M3Y)}

{\small * NPNTS = no. of integration points in radial coordinate (
$<$ 10000)}

{\small * RMAX = maximum radius size (%
$<$ 250 fm)}

{\small * NEPTS = number of points in energy ( $<$ 1000)}

{\small * IOPT \ \hspace{0.09in} NPNTS \ \hspace{0.09in}RMAX \
\hspace {0.09in} NEPTS}

{\small \ \hspace{0.09in} 1 \ \hspace{0.09in}9999 \ \hspace{0.09in}250.
\ \hspace{0.09in} 200}

{\small * V0 = depth of central potential }

{\small * VS0 = depth of spin-orbit potential }

{\small * R0 = radius of the potential }

{\small * AA = diffuseness of the potential }

{\small * RS0 = radius of the spin-orbit potential }

{\small * AAS = diffuseness of the spin-orbit potential }

{\small * RC = Coulomb radius (usually RC = R0) }

{\small
*----------------------------------------------------------------}

{\small * WS = V\_0 f(r,R0,AA) - V\_S0 (l.s) (r\_0\symbol{94}2/r)
d/dr f(r,RS0,AAS) }

{\small * f(r,R0,A) = [ 1 + exp((r-R0)/a) ]\symbol{94}(-1)}

{\small * r\_0 = 1.4138 fm is the Compton wavelength of the pion.
}

{\small
*----------------------------------------------------------------}

{\small * V0 \ \hspace{0.09in}R0 \ \hspace{0.09in}AA \
\hspace{0.09in}VS0 \ \hspace{0.09in} RS0 \ \hspace{0.09in} AAS \
\hspace{0.09in} RC}

{\small \ \hspace{0.09in} -42.3 \ \hspace{0.09in} 2.391 \ \hspace{0.09in} 0.52
\ \hspace{0.09in} -9.8 \ \hspace{0.09in} 2.391 \ \hspace{0.09in} 0.52
\ \hspace{0.09in} 2.391 }

{\small * If IOPT = 2, or else (but not 1), enter FC, FSO and RC:}

{\small * (in this case, insert a '*' sign in above row, or delete
it)}

{\small * FC = multiplicative factor of central part of M3Y
potential}

{\small * FSO = multiplicative factor of spin-orbit part of M3Y
potential}

{\small * RC = Coulomb radius }

{\small * FC \ \hspace{0.09in}FSO \ \hspace{0.09in} RC}

{\small * 1.5 \ \hspace{0.09in}0.2 \ \hspace{0.09in} 2.391}

{\small * Z1, Z2 = charges of the nuclei}

{\small * A1, A2 = masses of the nuclei (in nucleon mass units) }

{\small * Z1 \ \hspace{0.09in}A1 \ \hspace{0.09in} Z2 \
\hspace{0.09in} A2}

{\small \ \hspace{0.09in} 1. \ \hspace{0.09in}1. \ \hspace{0.09in} 4.
\ \hspace{0.09in} 7.}

{\small * EI, EF = initial energy, final energy}

{\small * L, J = orbital angular momentum, angular momentum j
(l+s) }

{\small * EI \ \hspace{0.09in}EF \ \hspace{0.09in} L \
\hspace{0.09in} J}

{\small \ \hspace{0.09in} 0. \ \hspace{0.09in}3. \ \hspace{0.09in} 1
\ \hspace{0.09in}1.5 }

{\small **********************************************}%

A run with this input file will show the presence of a sharp resonance at 631
keV with a width of approximately 50 keV. As for the case of bound states, the
same resonance can be obtained with a different set of WS potential
parameters, e.g. with the parameters shown below.

{\small * V0 \ \hspace{0.09in} R0 \ \hspace{0.09in} AA \
\hspace{0.09in} VS0 \ \hspace{0.09in} RS0 \ \hspace{0.09in} AAS \
\hspace{0.09in} RC}

{\small \ \hspace{0.09in}-28.65 \ \hspace{0.09in} 2.95 \ \hspace{0.09in} 0.52
\ \hspace{0.09in} -8.5 \ \hspace{0.09in} 2.95 \ \hspace{0.09in} 0.52
\ \hspace{0.09in} 2.95 }

The continuum states are calculated by the subroutine CONTWAVE and the Coulomb
wavefunctions are calculated by the subroutine COULOMB.

The \ phase-shifts and their derivatives\ with respect to energy are printed
in the output file CONT.TXT. The Figure\  3 shows these quantities for the
test case above.

\begin{figure}
[ptb]
\begin{center}
\includegraphics[
height=2.9421in,
width=3.3909in
]%
{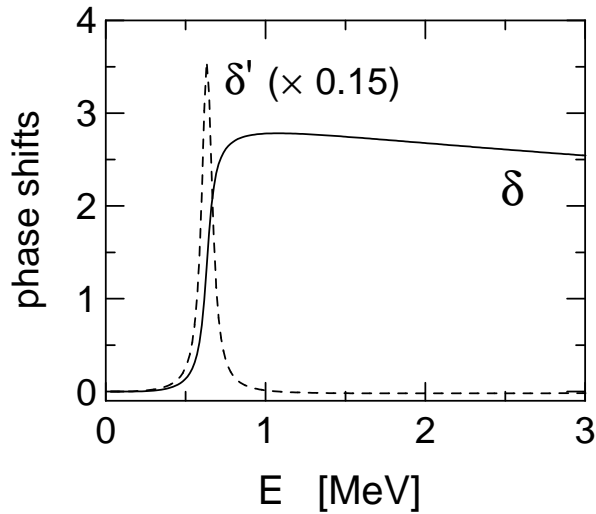}%
\caption{Phase-shift (solid line) and its derivative (dashed line)
for the p$+^{7}$Be system with the potential parameters described
in the text. The
1$^{+}$ resonance at 630 keV is observed. }%
\end{center}
\end{figure}

The program also allows for the output of the continuum wavefunction for a
given energy. The output of the wavefunction is printed in CWAVE.TXT. The real
part of the continuum wave function of the 1$^{+}$ resonance state at 630 keV
is shown in Figure\  2 (dotted line).

\subsection{Direct capture cross sections}

By choosing the option 5 the program calculates the direct capture cross
sections and related quantities. The input file is DICAP.INP. The calculations
are done in the subroutine DICAP. The output files are DICAP.TXT \ where the
strength functions (in units of e$^{2}$ fm$^{2\lambda}$), photodissociation
cross sections (in mb), direct capture cross sections (in mb), and the
astrophysical S-factors (in eV.b) are printed; DICAPL.TXT \ where the same
output is printed, prepared for using in a plot program; and SFAC.TXT where
the S-factor and its first and second derivatives with respect to the energy
are printed. These can be used in the calculation of the reaction rates by
using Eqs. \ref{sexp}, \ref{svap} and \ref{svap2}.

An input example is presented below.

{\small **********************************************}%

{\small * ******** Input of program DICAP *******}

{\small * IOPT = option for potentials: 1 (2) for Woods-Saxon
(M3Y)}

{\small * NPNTS = no. of integration points in radial coordinate (
$<$ 10000)}

{\small * RMAX = maximum radius size (%
$<$ 250 fm).}

{\small * NEPTS = number of points in energy ( $<$ 1000)}

{\small * IOPT \ \hspace{0.09in} NPNTS \ \hspace{0.09in} RMAX \
\hspace {0.09in} NEPTS}

{\small \ \hspace{0.09in}1 \ \hspace{0.09in}9999 \ \hspace{0.09in}250.
\ \hspace{0.09in}200}

{\small * N\_0 = nodes of the ground state wave function }

{\small * AIA = spin of the particle A (core)}

{\small * AIB = intrinsic spin of the particle B }

{\small * AIC = total angular momentum of the ground state of C =
A + B}

{\small * (channel spin)}

{\small * J0 = single-particle angular momentum }

{\small * L0 = orbital angular momentum}

{\small * EBOUND = binding energy of the ground state (absolute
value)}

{\small * N\_0 \ \hspace{0.09in} AIA \ \hspace{0.09in} AIB \
\hspace{0.09in} AIC \ \hspace{0.09in}AJ0 \ \hspace{0.09in} L0 \
\hspace{0.09in}EBOUND}

{\small \ \hspace{0.09in}0 \ \hspace{0.09in}1.5 \ \hspace{0.09in} 0.5
\ \hspace{0.09in} 2 \ \hspace{0.09in}1.5 \ \hspace{0.09in} 1 \ \hspace{0.09in}
0.14}

{\small * JOPT = 1 (0) if final state angular momentum, AICF, is
(is not) to be}

{\small * summed over all possible values. If JOPT=1, AICF can be
}

{\small * entered as any value.}

{\small * AICF = spin of the excited state after all angular
momentum coupling}

{\small * (channel spin)}

{\small * JOPT \ \hspace{0.09in} AICF}

{\small \ \hspace{0.09in} 1 \ \hspace{0.09in}1.}

{\small * Z1, Z2 = charges of the nuclei}

{\small * A1, A2 = masses of the nuclei (in nucleon mass units) }

{\small * Z1 \ \hspace{0.09in} A1 \ \hspace{0.09in} Z2 \
\hspace{0.09in} A2}

{\small \ \hspace{0.09in}1. \ \hspace{0.09in} 1. \ \hspace{0.09in}4.
\ \hspace{0.09in}7. }

{\small * V0 = depth of central potential }

{\small * R0 = radius of the central potential }

{\small * AA = diffuseness of the central potential }

{\small * VS0 = depth of spin-orbit potential }

{\small * RS0 = radius of the spin-orbit potential }

{\small * AAS = diffuseness of the spin-orbit potential }

{\small * RC = Coulomb radius (usually, RC = R0) }

{\small
*-----------------------------------------------------------------}

{\small * WS = V\_0 f(r,R0,AA) - V\_S0 (l.s) (r\_0\symbol{94}2/r)
d/dr f(r,RS0,AAS) }

{\small * f(r,R0,a) = [ 1 + exp((r-R\_0)/a) ]\symbol{94}(-1)}

{\small * r\_0 = 1.4138 fm is the Compton wavelength of the pion.
}

{\small
*-----------------------------------------------------------------}

{\small * V0 \ \hspace{0.09in}R0 \ \hspace{0.09in}AA \
\hspace{0.09in}VS0 \ \hspace{0.09in}RS0 \ \hspace{0.09in}AAS \
\hspace{0.09in}RC}

{\small \ \hspace{0.09in} -44.658 \ \hspace{0.09in} 2.391 \ \hspace{0.09in}
0.52 \ \hspace{0.09in} -9.8 \ \hspace{0.09in} 2.391 \ \hspace{0.09in} 0.52
\ \hspace{0.09in} 2.391 }

{\small * If IOPT = 2, or else (but not 1), enter FC, FSO and RC:}

{\small * (in this case, insert a '*' sign in above row, or delete
it)}

{\small * FC = multiplicative factor of central part of M3Y
potential}

{\small * FSO = multiplicative factor of spin-orbit part of M3Y
potential}

{\small * RC = Coulomb radius }

{\small * FC \ \hspace{0.09in} FSO \ \hspace{0.09in} RC}

{\small * \ \hspace{0.09in}1.5 \ \hspace{0.09in}0.2 \
\hspace{0.09in} 2.391}

{\small * EI,EF = initial relative energy, final relative energy}

{\small * EI \ \hspace{0.09in}EF }

{\small \ \hspace{0.09in}0. \ \hspace{0.09in}3. }

{\small * NS1,NP1,NP3,ND3,ND5,NF5,NF7 = (1) [0] for inclusion (no
inclusion) }

{\small * of s1/2, p1/2, p3/2, d3/2, d5/2, f5/2, and f7/2 partial
waves}

{\small * NS \ \hspace{0.09in} NP1 \ \hspace{0.09in} NP3 \
\hspace{0.09in} ND3 \ \hspace{0.09in} ND5 \ \hspace{0.09in} NF5 \
\hspace{0.09in} NF7}

{\small \ \hspace{0.09in} 1 \ \hspace{0.09in} 0 \ \hspace{0.09in} 0
\ \hspace{0.09in} 1 \ \hspace{0.09in} 1 \ \hspace{0.09in} 0 \ \hspace{0.09in}
0}

{\small * MP = multipolarity: 0 (M1), 1 (E1), 2 (E2) }

{\small * SF = Spectroscopic factor }

{\small * MP \ \hspace{0.09in} SF}

{\small \ \hspace{0.09in} 1 \ \hspace{0.09in}1.}

{\small * GA = magnetic moment (in units of the nuclear magneton)
of }

{\small * particle A (core)}

{\small * GB = magnetic moment of particle B (proton, neutron,
alpha, etc.)}

{\small * GA \ \hspace{0.09in}GB}

{\small \ \hspace{0.09in} -1.7 \ \hspace{0.09in} 5.58}

{\small **********************************************}%

Only the results for the astrophysical S-factor, S$_{17}$, for the reaction
p$+^{8}$B will be shown. They are plotted in Figure\  4, together with the
experimental data of several experiments \cite{MSU,GSI1,GSI2,Weiz}. The first
three set of data (MSU, GSI-1, and GSI-2) were obtained by using the Coulomb
dissociation method \cite{BBR86}. The other experimental results \cite{Weiz}
were obtained via a direct measurement. The dashed line shows the result of
the calculated S-factor, obtained with the bound state wavefunction calculated
with the same Woods-Saxon parameters as in the above input file. \ The dashed
line represents the S-factor one obtains by changing the bound and continuum
states using another set of Woods-Saxon potential parameters, which yields the
same binding for $^{8}$B, namely:

{\small * V0 \ \hspace{0.09in} R0 \ \hspace{0.09in} AA \
\hspace{0.09in} VS0 \ \hspace{0.09in} RS0 \ \hspace{0.09in} AAS \
\hspace{0.09in} RC}

{\small -30.55 \ \hspace{0.09in} 2.95 \ \hspace{0.09in} 0.52 \ \hspace{0.09in}
-8.53 \ \hspace{0.09in} 2.95 \ \hspace{0.09in} 0.52 \ \hspace{0.09in} 2.95 }
\begin{figure}
[ptb]
\begin{center}
\includegraphics[
height=3.4566in,
width=4.4719in
]%
{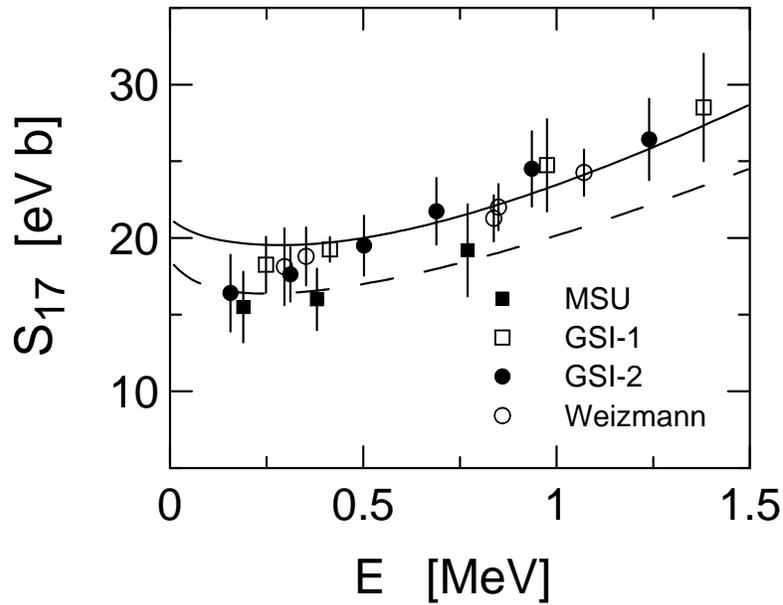}%
\caption{Astrophysical S-factor for the reaction p$+^{8}$B. The data are the
experimental points of recent experiments \cite{MSU,GSI1,GSI2,Weiz}. The solid
and dashed curves are results of calculations with different choices of the
Woods-Saxon potential which reproduces the binding of $^{8}$B.}%
\end{center}
\end{figure}

\section{Things to do}

1 - Use the input files described above and reproduce the Figures
1-4.

2 - Try to reproduce some of the radiative capture cross sections
presented in the compilation of Ref. \cite{Angulo}.

3 - Show that for neutron halo nuclei the radiative capture cross
sections follow the dependence described by equations \ref{qqq1}
and \ref{qqq2}.

\section{Acknowledgments}

I am indebted to Sam Austin and Horst Lenske for many useful
discussions. This material is based on work supported by the
National Science Foundation under Grants No. PHY-0110253,
PHY-9875122, PHY-007091 and PHY-0070818.

\end{document}